\documentclass[aps,prl,twocolumn,superscriptaddress,floatfix]{revtex4-1}
\usepackage{graphicx,amsmath,subfigure,epsfig,psfrag,verbatim}
\begin{document}
\title{Random Field Driven Spatial Complexity at the Mott Transition in VO$_2$}

\author{Shuo Liu}
\email{liu305@purdue.edu}
\affiliation{Department of Physics, Purdue
University, West Lafayette, IN 47907, USA}
\author{B. Phillabaum}
\affiliation{Department of Physics, Purdue University, West
Lafayette, IN 47907, USA}
\author{E.~W. Carlson}
\affiliation{Department of Physics, Purdue University, West
Lafayette, IN 47907, USA}
\author{K.~A. Dahmen}
\affiliation{Department of Physics, University of Illinois,
Urbana-Champaign, IL 61801, USA}
\author{N.~S. Vidhyadhiraja}
\affiliation{Jawaharlal Nehru Centre for Advanced Scientific
Research, Bangalore 560064, India}
\author{M.~M. Qazilbash}
\affiliation{Department of Physics, College of William and Mary,
Williamsburg, VA 23187, USA}
\author{D.~N. Basov}
\affiliation{Department of Physics, University of California-San
Diego, La Jolla, CA 92093, USA}
\date{\today}

\begin{abstract}
We report the first application of critical cluster techniques to the Mott
metal-insulator transition in vanadium dioxide. We show that the
geometric properties of the metallic and insulating puddles observed
by scanning near-field infrared microscopy are consistent with the
system passing near criticality of the random field Ising model as
temperature is varied.  The resulting large barriers to equilibrium
may be the source of the unusually robust hysteresis phenomena
associated with the metal-insulator transition in this system.
\end{abstract}

\maketitle

%{\bf Introduction}.
The Mott metal-insulator transition in VO$_2$ has the potential to
produce disruptive technologies, such as memristors, memory
capacitors, ultrafast switches, and possibly even neuromorphic
circuits \cite{yang-2011}. Yet despite decades of research into the
Mott metal-insulator transition in VO$_2$, the nature of the phase
transition, and in particular the broad hysteretic behavior
accompanying it, is not yet understood. The Mott metal-insulator
transition has some universal features of the liquid-gas transition
\cite{limelette-science,papaniko}: the transition occurs through a
first order phase transition line broadened by conductivity
hysteresis, which terminates at a classical critical point
associated with general Ising universality
\cite{limelette-science,papaniko,Kagawa,Castellani,Landau1}. In
addition, in VO$_2$ a percolation model has been proposed through
the study of resistance avalanches \cite{sharoni-prl}, while in
V$_2$O$_3$ the clean 3D Ising model has been implicated
\cite{limelette-science}. By applying quantitative critical cluster
techniques to study the criticality, we show that a key ingredient
missing from prior treatments is the prominent role of disorder,
with important implications for the robust hysteresis effects
associated with the metal-insulator transition.

\begin{figure}
\centering
\includegraphics[width=0.95\columnwidth]{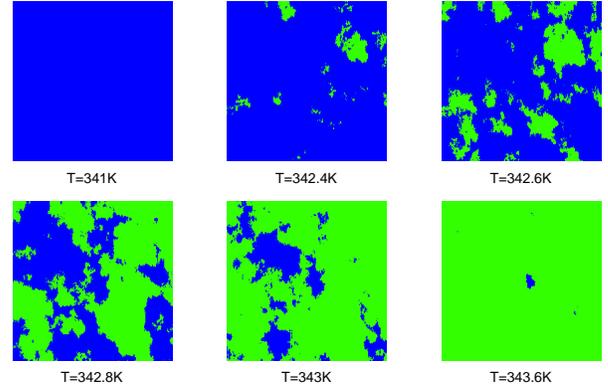}%
\caption{Scanning Near-field Infrared Microscopy on VO$_2$ as
temperature is increased through the Mott metal-insulator transition
regime, over the same $4\mu m \times 4\mu m$ sample area. These
figures are Ising-mapped from the original SNIM images in Ref.
\cite{basov-science} with threshold scattering amplitude
$a_{th}=2.5$. The metallic regions, colored green, give higher
near-field scattering amplitude, $a>a_{th}$, compared with the
insulating regions with $a\le a_{th}$, colored blue.}
 \label{SNIM-isingmap}
\end{figure}

As shown in Fig.~\ref{SNIM-isingmap}, scanning near-field infrared
microscopy (SNIM) on VO$_2$ \cite{basov-science} reveals the complex
pattern formation associated with the transition from the low
temperature insulating phase to the high temperature metallic phase.
We apply recently developed cluster techniques
\cite{phillabaum-2012} to the observed multiscale patterns of
inhomogeneous local conductivity, revealing critical exponents such
as the fisher exponent $\tau$, volume fractal dimension $d_v$, and
hull fractal dimension $d_h$, both for clusters and avalanches. This
method has the advantage that it can access the true universality
class of the phase transition, unlike methods based on macroscopic
conductivity in which the exponents are affected by the contrast
ratio between metallic and insulating regions \cite{papaniko}. Our
analysis is the first to quantitatively incorporate disorder into
the study of the criticality, allowing us to determine the relative
importance of disorder and interactions, and identify the dominant
type of disorder.

%{\bf Model}.
Near a critical point, the correlation length grows to become the
dominant length scale, and it is possible to map the real physical
system to a coarse-grained model with the same universal features.
In the original SNIM data images \cite{basov-science}, we assign
Ising variable $\sigma = 1$ (metallic) or $-1$ (insulating) to each
coarse-grained region
\cite{limelette-science,papaniko,Kagawa,Castellani,Landau1}. The
threshold amplitude $a_{th}$ for identifying the metallic regions
(those with scattering amplitude $a > a_{th}$) and insulating
regions (with $a\le a_{th}$) is about 2.5 \cite{basov-science, SI},
which we use throughout this Letter. We furthermore incorporate
disorder into the model:
\begin{eqnarray}
H&=&-\sum^{\infty}_{\langle ij\rangle_\parallel}(J +\delta
J_{ij})\sigma_i \sigma_j -\sum^{L_z}_{\langle
ij\rangle_{\perp}}(J+\delta J_{ij})\sigma_i
\sigma_j \nonumber \\
&-&\sum_i(h+h_i)\sigma_i~, \label{eqn:model}
\end{eqnarray}
%EC 8/9/13 Define the sum
where the sum runs over the coarse-grained regions (sites)
consisting of a cubic lattice, chosen with spacing at least as small
as the resolution of the images to be studied. The tendency for
neighboring regions to be of like character is modeled as a nearest
neighbor ferromagnetic interaction $J>0$. Because the data
considered is that of a thin film, the sum over Ising variables in
the plane of the film (denoted by $\parallel$) extends to infinity,
but the sum over Ising variables perpendicular to the film
surface(denoted by $\perp$) is finite, confined by the film
thickness $L_z$. Depending on the size of the correlation volume
compared to the size of the system, the sample may display
two-dimensional (2D) or three-dimensional (3D) critical behavior
near criticality.

\begin{figure}
  \centering
  \includegraphics[width=0.95\columnwidth]{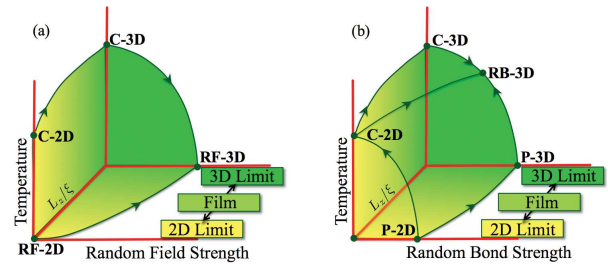}
  \caption{
Schematic equilibrium phase diagrams and fixed points of
Eqn.~\ref{eqn:model}. Two classes of disorder: (a) Random field
disorder (with or without random bond disorder); (b) Random bond
disorder in the absence of random field disorder. Solid regions
denote the ordered phase, from 2D (yellow region, applicable when
the correlation length $\xi \gg L_z$) to 3D (green region,
applicable for $\xi$ large but less than $L_z$). Solid green lines
represent continuous phase transitions. The green arrow(flow) on
each phase transition line points to the solid green circle
representing the fixed point controlling the long distance
(universal) behavior of that phase transition.
  }
  \label{pd-3D}
\end{figure}

At the order parameter level, there are two broad classes of
disorder: local energy density disorder (which we incorporate as
random bond disorder), and random field disorder \cite{cardy-book}.
Random bond disorder is included through the term $\delta J_{ij}$,
and $h_i$ represents random field disorder, which is chosen from a
gaussian probability distribution centered about zero, with variance
$R$. $R$ is often called the disorder parameter or just disorder.
The field $h$ represents a generalized external field which couples
with the local Ising variables. Figure \ref{pd-3D} shows the
schematic phase diagrams associated with this general Ising model of
Eqn.~\ref{eqn:model}, emcompassing several universality classes both
in 2D and 3D. Note that random field disorder is always relevant in
the renormalization group sense, and in fact if both random bond and
random field disorder are present, the associated stable fixed
points are those of the random field disorder.
%Figure \ref{pd-2D}(a) shows a cross section of the phase diagram in
%the absence of quenched disorder, and
%%EC-04/18/14 order of RB, RF
%Fig.~\ref{pd-2D}(b) and
%\ref{pd-2D}(c) show cross sections of the phase diagram at finite
%random field disorder and at finite random bond disorder,
%respectively. The
%blue dotted line is the temperature $T_p$ at which minority spin
%clusters first percolate.
%Notice that this happens {\em inside} the ordered phase $T_p < T_c$
%in the absence of random field disorder \cite{C3Dfrac,RB3D,nofrac},
%but that it coincides with the phase transition in the presence of
%random field disorder, $T_p = T_c$
%\cite{Coniglio:1980p979,Middleton:2002p986}.

Using our mapping to Ising variables (Fig.~\ref{SNIM-isingmap}), we
track the geometric clusters, defined as connected sets of nearest
neighbor sites (pixels) with the same color. We then use the
statistics of the sizes and shapes of these geometric clusters to
identify the cause of the complex pattern formation. In comparing
the spatial complexity revealed in the SNIM data to theory, there
are 7 fixed points to consider, as shown in Fig.~\ref{pd-3D}. In the
two-dimensional case, these consist of the clean Ising model (C-2D,
{\em i.e.} in the absence of any material disorder), uncorrelated
percolation (P-2D), and the random field Ising model (RF-2D). Note
that random bond disorder is irrelevant in the renormalization group
sense in 2D, and the phase transition is therefore governed by clean
Ising model exponents set by the C-2D fixed point. Although the
RF-2D fixed point is unstable, its critical behavior can be observed
for weak enough finite disorder. In the three-dimensional case, the
possible fixed points are the clean Ising model (C-3D), uncorrelated
percolation (P-3D), the random bond Ising model (RB-3D), and the
random field Ising model (RF-3D). For the exponents extraction, in
order to reduce noise due to the finite field of view (FOV), we
apply the logarithmic binning method throughout this Letter, which
is a standard technique for analyzing power law behavior
\cite{Newman}. The value of critical exponents are obtained by
taking the discrete logarithmic derivative (DLD)
\cite{Middleton:2002p986}.

%
%\begin{figure}
%  \centering
%  \includegraphics[width=0.95\columnwidth]{figs/PhaseDiagram-Ising-CrossSections-Percolation.eps}%
%  \caption{The Temperature--$L_z/\xi$ plane cross sections of
%  Fig.~\ref{pd-3D}.
%  (a) Clean Ising Model.
%  %EC-04/18/14 order of RB, RF
%  (b) Ising model in the presence of finite random field disorder, with or without random bond disorder.
%  (c) Ising model in the presence of finite random bond disorder
%  and no random field disorder.
%  $T_p$ is the percolation temperature, and $T_c$ is the critical temperature.}
%  \label{pd-2D}
%\end{figure}

%{\bf Results.}

From the SNIM images, we extract three critical exponents, which are
the Fisher exponent $\tau$, the volume fractal dimension $d_v$, and
the hull fractal dimension $d_h$, as entailed by the self-similarity
of the geometric clusters near certain critical points. $\tau$
characterizes the cluster-size distribution $D(s)$, which is the
histogram of cluster sizes $s$, and scales as $D(s) \propto
s^{-\tau}$. $d_v$ and $d_h$ characterize the fractal nature of
cluster sizes $s$ and hulls $h$. For the cluster sizes we have $s
\propto R_s^{d_v}$, where $R_s$ is the radius of gyration of the
cluster \cite{Percobook}. For the cluster surfaces, we have $h
\propto R_h^{d_h}$. $R_h$ here refers to the radius of gyration of
{\em all} the sites enclosed by the hull, including any subclusters
inside. Throughout this letter, we analyze the power law behaviors
using only the internal clusters, i.e., clusters which do not
intersect the boundary of the FOV, in order to mitigate finite FOV
effect to the extraction of exponents. Although from our analysis
$d_v$ and $d_h$ do not appear to be affected by this effect,
estimates of $\tau$ in a finite FOV are always skewed to lower
values because of a pronounced bump in the scaling function of the
cluster size distribution \cite{dahmen-perkovic, sethna-window,SI}.
Within a cutoff in the decades of scaling, $D(s)$ of internal
clusters would share the same power law as the full system
\cite{sethna-window}, therefore $\tau$ could be extracted more
accurately.

\begin{figure}
\centering
\includegraphics[width=0.95\columnwidth]{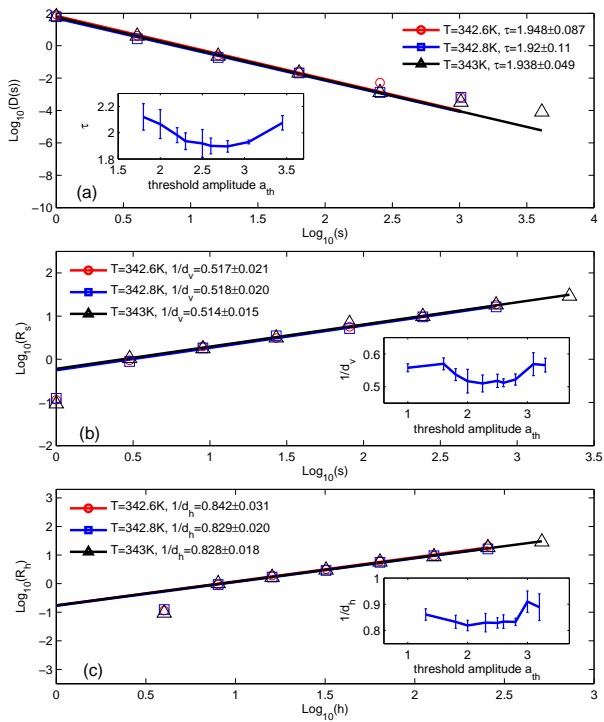}
\caption{Power law fits of internal clusters from the VO$_2$ SNIM
imaging datasets with threshold scattering amplitude $a_{th}=2.5$
for (a) $D(s) \propto s^{-\tau}$, (b) $R_s \propto s^{1/d_v}$, and
(c) $R_h \propto h^{1/d_h}$, at three intermediate temperatures
$T=342.6K$, $T=342.8K$, and $T=343K$. The insets show $\tau$,
$1/d_v$, and $1/d_h$ (with error bars) extracted using the same
method as a function of $a_{th}$ at T=342.8K.} \label{cluster}
\end{figure}

Figure \ref{cluster} shows the extraction of $\tau$, $d_v$, and
$d_h$ by applying our cluster techniques to the finite size SNIM
image data through the Mott transition using internal clusters. The
main panels show explicitly the power law fits of these critical
cluster exponents at three intermediate temperatures (which have
enough clusters for good statistics) with threshold scattering
amplitude $a_{th}=2.5$. For $\tau$, two decades of scaling are
evident, and the remaining points fall off the scaling regime due to
the use of internal clusters \cite{sethna-window}; and for the
fractal dimensions, with the first bin (which contains $s=1$
clusters) excluded \cite{SI}, robust power law scaling extends over
multiple decades, encompassing the entire FOV. The robust power law
behaviors as well as the fact that the values of the cluster
exponents are the same within error bars for different intermediate
temperatures corroborates the idea that the system is near some
critical end point of the Mott transition. The insets of
Fig.~\ref{cluster} show our extracted $\tau$, $d_v$, and $d_h$ using
different $a_{th}$ at the representative temperature $T=342.8K$ which
is closest to criticality, and within error bars these critical
exponents are robust against changes in threshold amplitude within a
broad stable region around $a_{th}=2.5$, consistent with that given
by the definition in the experiment paper \cite{basov-science, SI}.
This independence of results with respect to microscopic details
also reflects universal behavior near criticality. Using
$a_{th}=2.5$ and T=342.8K as representative conditions, we have
$\tau=1.92\pm0.11$, $d_v=1.93\pm0.07$, and $d_h=1.21\pm0.03$ for the
spatially complex clusters near the critical end point of the Mott
transition in VO$_2$.

%Due to the limitations of real experimental data (finite FOV, few
%data configurations), for clearness purpose it is reasonable to use
%the DLD result under $a_{th}=2.5$ and T=342.8K to represent the
%universal data-extracted values of the cluster exponents in the
%whole Mott transition regime.

\begin{figure}
\centering
\includegraphics[width=0.95\columnwidth]{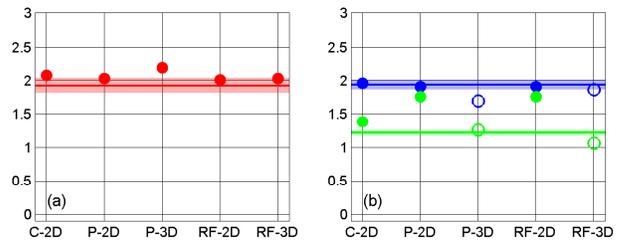}
\label{chart} \caption{Exponent comparison charts. (a) Critical
exponent $\tau$. (b) Critical exponents $d_v$ (blue) and $d_h$
(green). The horizontal lines are our extracted exponents from the
SNIM images, with the shaded regions being their error bars. The
circles represent theoretical values \cite{phillabaum-2012, SI} for
the fixed points of Eqn.~\ref{eqn:model}. When comparing with 3D
models, we have to use the effective values of $d_v^*$ and $d_h^*$
corresponding to taking 2D cross sections of the clusters embedded
in 3D, {\em i.e.} $d_v^* =2d_v/3$ and $d_h^* = d_h/2$
\cite{phillabaum-2012, SI}. These effective values are represented
by the open circles.} \label{fig:chart}
\end{figure}

%{\bf Discussion}
%With three critical exponents in hand ($\tau$, $d_v$, and $d_h$), we
%can now analyze whether the results are consistent with any of the
%fixed points associated with the general disordered Ising model of
%Eqn.~\ref{eqn:model}. If so, this then identifies the universality
%class of the Mott critical point.

Before comparing the numerical values of the critical exponents
revealed by our analysis of the data to the known theoretical values
of each fixed point, the C-3D and RB-3D fixed points can be ruled
out {\em a priori}. Since geometric clusters are not fractal at
these two fixed points \cite{C3Dfrac,RB3D,nofrac}, they cannot be
the cause of the robust power law behavior observed in
Figs.~\ref{cluster}(b) and \ref{cluster}(c). Five fixed points
remain to be considered (C-2D, RF-2D, RF-3D, P-2D, and P-3D), and
the geometric clusters do exhibit a fractal nature at these fixed
points \cite{Percobook, Middleton:2002p986, RF2Ddv, C3Dfrac}. In the
slab geometry considered here (and also for layered materials), it
is also necessary to consider possible dimensional crossovers.
Within this set of candidate fixed points, there are two possible
dimensional crossovers. In the presence of random field disorder,
the dimensional crossover is between RF-2D and RF-3D. (For a slab
geometry, exponents will drift from the 3D fixed point toward the
associated 2D fixed point, once the correlation length in the c-axis
direction begins to exceed the slab thickness.) The other possible
dimensional crossover is between C-2D and the 3D percolation points
$T_p < T_c$ in the clean and random bond Ising models. Although the
exponents we require have not been explicitly reported in the
literature, the expectation in the literature is that the exponents
should follow those of uncorrelated percolation, P-3D
\cite{Cambier:1986tp,huse-leibler,nofrac}.
%Remember that
%in these two cases, at 3D, we have $T_p < T_c$, and the geometric
%clusters go fractal inside the ordered phase.

As shown in Fig.~\ref{fig:chart}, the values of $\tau$ and $d_v$
resulting from our analysis of the data are fairly close to all
fixed points other than P-3D. Therefore, uncorrelated 3D percolation
can be ruled out as an origin of the power law behavior of the
statistics and geometry of the metal and insulator islands. Note,
however, that the theoretical values of the remaining four fixed
points are all fairly close in value for these two exponents, making
it hard to distinguish among the remaining fixed points using only
these two measures. On the other hand, $d_h$ varies significantly
for the various fixed points. We see immediately from the comparison
chart that the theoretical values of $d_h$ at P-2D and RF-2D are
inconsistent with the data. Whereas the closest match for this
measure is P-3D, that is inconsistent with the other two critical
exponents, and must remain ruled out.

Only two candidate fixed points remain (C-2D and RF-3D), but in
Fig.~\ref{fig:chart}(b), both show about a $13\%$ discrepancy
between the data and the theoretical model for the value of $d_h$,
which is significantly larger than the typical error of about $3\%$
of the exponents extracted from the data. We turn then to the
possibility of dimensional crossovers. In the case of the clean or
random bond Ising model, power law behavior of geometric clusters
will drift from P-3D to C-2D as larger length scales are observed.
However, the value of $d_v$ is much closer to C-2D than to P-3D,
consistent with the geometric clusters being near their 2D limit,
whereas the value of $d_h$ is much closer to P-3D than to C-2D,
which would indicate that the geometric clusters are near their 3D
limit. This inconsistency strongly argues against the P-3D to C-2D
dimensional crossover as the origin of the power law behavior.

The other candidate dimensional crossover is from RF-3D to RF-2D.
The exponent comparisons show no inconsistency with this
explanation, since within error bars $\tau$ and $d_v$ match the
whole RF-2D to RF-3D crossover regime, and $d_h$ is also consistent
with this dimensional crossover. Therefore, the random field Ising
universality class best describes the critical behavior of the Mott
transition in the VO$_2$ thin film, revealing the key role played by
disorder effects in these systems. Observations at longer length
scales can be used to test this hypothesis: if our identification is
correct, then the values of all exponents should drift toward RF-2D
with larger fields of view in a film geometry such as the present
one.

\begin{figure}
\centering
\includegraphics[width=0.95\columnwidth]{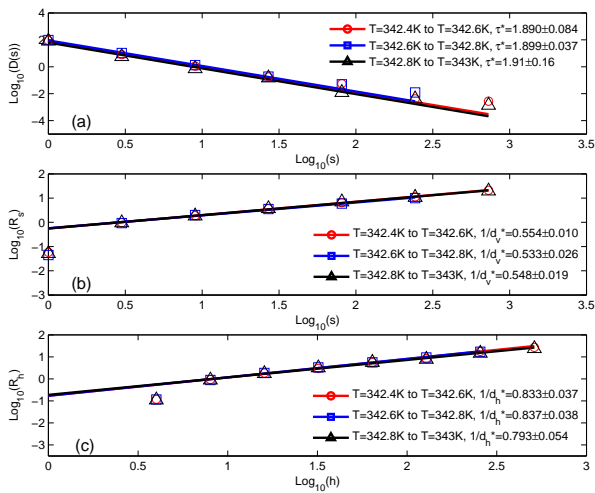}
\caption{Power law fits of internal avalanches from the VO$_2$ SNIM
imaging datasets with threshold scattering amplitude $a_{th}=2.5$
for (a) $D(s) \propto s^{-\tau^*}$, (b) $R_s \propto s^{1/d_v^*}$,
and (c) $R_h \propto h^{1/d_h^*}$, at three intermediate intervals
with $\Delta T= 0.2K$ through the Mott transition}.
\label{avalanche}
\end{figure}

In order to further test whether the spatial complexity is driven by
random field effects, we also analyze the avalanches which, in the
context of the VO$_2$ experiment, are defined as the difference
between Ising maps between two neighboring temperatures. We study
the critical avalanche exponents arising from the self-similarity of
avalanches and are directly extractable from the experimental image
datasets. Figure \ref{avalanche} shows the power law fits of the
internal avalanches with threshold $a_{th}=2.5$ for the
data-extracted $\tau^*$, $d_v^*$, and $d_h^*$, using the same
techniques as for the clusters in Fig.~\ref{cluster}. The star
symbol is used to denote that these exponents are for avalanches.
Similar to their cluster counterparts, the cut-off for internal
avalanche scaling for $\tau^*$ is about 1.5 to 2 decades of scaling,
and $d_v^*$ and $d_h^*$ show multiple decades of power law scaling
behavior. For different intervals, the extracted critical avalanche
exponents are consistent with each other within error bars. This
data-extracted robust universal power law scaling of avalanches
provides further evidence that the random field universality is
underlying the complex pattern formation at the Mott transition in
the VO$_2$ thin film, since non-trivial scaling behavior of
avalanches is characteristic of random field physics.

Since the avalanches are taken from the finite interval between
scanned images, the data-extracted $\tau^*$ in our context
characterizes the scaling of the field integrated avalanche size
distribution $D_{int}(s)$ near criticality. The corresponding
exponent is reported to be $2.03\pm0.03$ for RF-3D
\cite{dahmen-perkovic,kd1, kd2} and $1.3\pm0.1$
\cite{PhysRevE.62.7470} for RF-2D, and the corresponding
data-extracted one from Fig.~\ref{avalanche}(a) lies in between,
consistent with the conclusion of a dimensional crossover from RF-3D
to RF-2D. With $d_v^*$ and $d_h^*$ only reported for RF-3D
\cite{kd2}, our data-extracted fractal dimensions
Fig.~\ref{avalanche}(b) and (c) could not be directly compared with
the theoretical ones. However, by comparison with
Fig.~\ref{cluster}(b) and (c), within error bars $d_v$ and $d_h$
extracted from avalanches have the same values as those from
clusters, consistent with Ref.~\cite{kd2}, which show that clusters
and avalanches have the same exponents in random field Ising model
(RFIM).

The dominance of random field disorder is quite reasonable in a real
physical system, arising from defects and impurities in the crystal
structure; randomness in grain size, orientation, and boundary
structure; and imperfections in the substrate. In particular, it has
been shown that the temperature at which grains transition depends
on grain size \cite{doi:10.1021/nl8031839}. In addition, we have
previously shown that metallicity nucleates first near the grain
boundaries (see Fig. 6 of Ref~\cite{PhysRevB.80.115115}). Since
physical temperature maps to effective field in our model, these
effects map to random field disorder in the model. (See SI for a
more thorough discussion.) One prediction of this line of reasoning
is that grain size, since it affects the random field strength in
the model, is expected to correlate with the hysteresis width.

Several other characteristics of the data serve to corroborate the
random field hypothesis. For example, as the Mott transition
proceeds, metallic puddle formation occurs via nucleation and not
exclusively by front propagation \cite{crackling-nature}, as can be
seen in Fig.~\ref{SNIM-isingmap}. In addition, as temperature
increases through the transition (analogous here to sweeping the
generalized field $h$ of Eqn.~\ref{eqn:model}), domains change from
$\sigma = -1$ to $\sigma = +1$, but never revert. This no-passing
rule is consistent with the random field universality class
\cite{kd2,PhysRevLett.70.3347,PhysRevLett.68.670}, indicating that
the effects of quenched disorder dominate over thermal fluctuations.
The dominance of quenched disorder over thermal effects is further
corroborated by the reproducibility of the SNIM images, in that
repeated near-field scans in the insulator-to-metal transition
regime over the same sample area and at a fixed temperature show
nearly identical patterns of metallic puddles in the insulating host
\cite{basov-science, Qazilbash:2009p1063}.

Finally, perhaps the most tell-tale sign of random field behavior in
the data is the large width of the hysteresis in temperature, about
$7.5K$ for temperature sweeps taking several minutes in a thin film
\cite{Qazilbash:2009p1063,Kumar:2013p1105}. The extreme critical
slowing down of the random field case
%makes these timescales far larger in the present case:
%because the random field fixed point is a zero temperature fixed point,
means that barriers to equilibration grow much more severely as
criticality is approached than the usual ``critical slowing down''
would predict. Hysteresis typically becomes long-lived near the
critical endpoint of a first order phase transition, with the
relaxation time $t_{rel}$ diverging as a power law near the critical
endpoint \cite{binder-nucleation}, $t_{rel} \propto \xi^z \propto
|g-g_c|^{- \nu z}$, where $g$ represents the relevant variable,
whether temperature $T$ or disorder strength $R$. $\xi$ is the
correlation length, $\nu$ is the exponent of the correlation length,
and $z$ is the dynamical exponent.
%with the nucleation rate $1/\tau_{\rm nuc} \propto \omega_o^{(-\beta \delta + \beta + \nu z)}$
%diverging as a power law as the critical endpoint is approached,
%where $\omega_o$ is the attempt frequency.\cite{binder-nucleation}
Rather than a mere power law, barriers to equilibration grow
exponentially near the RFIM critical endpoint
\cite{fisher-exponential-barriers}, $t_{rel}^{\rm RFIM} \propto {\rm
exp}[\xi^{\theta}] \propto {\rm exp}[1/|g-g_c|^{\nu \theta}]$, where
$\theta$ is the "violation of hyperscaling" exponent. These
exponentially large barriers to equilibration are consistent with
the large width of hysteresis loops evident in this Mott
metal-insulator transition. In addition, the large critical region
typically associated with RFIM physics \cite{dahmen-perkovic} is
consistent with the wide range of parameters over which the broad
hysteresis is observed.

%{\bf Conclusions.}
In conclusion, by applying newly developed cluster techniques to
SNIM data on VO$_2$, we have shown using three different critical
exponents that the critical end point of the Mott transition is in
the universality class of the random field Ising model. This finding
reveals a delicate interplay between interaction and disorder in
these systems. The random field Ising universality class has also
recently been shown to account for the spatial structure of the
locally oriented domains observed via scanning tunneling microscopy
on cuprate superconductors \cite{phillabaum-2012,seamus-glass}. This
further emphasizes the important role played by disorder in strongly
correlated electron systems, and indicates that there may be
universality to the spatial complexity \cite{Dagotto,
dagotto-moreo-manganite} observed in a wide variety of strongly
correlated electron systems. The cluster techniques employed here
can readily be applied to 2D images in the context of other
materials and microscopy techniques for the study of critical
behavior.

\begin{acknowledgments}
We thank H.~Aubin, J.~Honig, and A.~Zimmers for helpful
conversations. S.L., B.P., and E.W.C. acknowledge support from NSF
Grant No. DMR 11-06187. E.W.C. acknowledges receipt of an APS-IUSSTF
Professorship Award, and thanks JNCASR for hospitality. K.A.D.
acknowledges support from NSF Grant No. DMR 10-05209 and NSF Grant
No. DMS 10-69224. N.S.V. acknowledges IUSSTF, JNCASR and Purdue
University for funding the visitor exchange program. M.M.Q.
acknowledges support from NSF Grant No. DMR 12-55156. D.N.B.
acknowledges support from DOE-BES.
\end{acknowledgments}

\bibliography{VOTref}

\end{document}